\documentclass[aps,preprint,showpacs,floats,epsf,epsfig,nofootinbib,12pt]{revtex4}
\usepackage{graphicx}
\usepackage{epsfig}
\usepackage[dvips]{color}

\def\beq{\begin{eqnarray}}
\def\eeq{\end{eqnarray}}

\def\ra{\rangle}

\begin{document}

\title{The radiative decays of $0^{++}$ and $1^{+-}$ heavy  mesons}

\vspace{1cm}

\author{ \footnote{Corresponding author}Hong-Wei Ke$^{1*}$\footnote{khw020056@hotmail.com},
        Xue-Qian Li$^2$\footnote{lixq@nankai.edu.cn} and
        Yan-Liang Shi$^{2*}$\footnote{shiyanliang08@gmail.com} }

\affiliation{  $^{1}$ School of Science, Tianjin University, Tianjin 300072, China \\
  $^{2}$ School of Physics, Nankai University, Tianjin 300071, China }

\vspace{12cm}

\begin{abstract}
The radiative decay is believed to be an ideal lab to study hadronic
structure of newly observed resonances because the reactions are
governed by only the electromagnetic interaction (tree level).
However, to obtain correct theoretical values, one has to properly
deal with the non-perturbative QCD effects in the wavefunction and
hadronization. In this work we derive the formulas for the radiative
decays of $0^{++}$ and $1^{-+}$ heavy mesons in the light front
quark model (LFQM). Because $\mathcal{B}(\chi_{c0}\rightarrow
J/\psi\gamma)$ is well measured, the theoretical evaluation of the
transition rate can be used to test our approach. Within this
theoretical framework, the width of $\chi_{b0}\rightarrow
\Upsilon(1S)\gamma$ is evaluated. The formulas can be applied to
identify the inner structures of new resonances, for example the
isospin of $h_{c(b)}$ and the structure of $D_s(2317)$, via
processes $h_c\rightarrow \eta_c\gamma$, $h_b\rightarrow
\eta_b\gamma$ and $D_s(2317)\rightarrow D_s^*+\gamma$.


\end{abstract}

\maketitle

\section{Introduction}
In the field of heavy hadrons, it is noticed that some experimental
observations obviously deviate from our theoretical predictions, so
definitely, such ``anomalies" need to be clarified. Supposing the
experimental measurements are right, there must be some loopholes in
our present understanding of the nature, either there exists a
contribution from new physics beyond the standard model (SM), or the
concerned hadrons have exotic structure such as hybrid, multi-quark
etc. It is believed that radiative decays on this aspect provide an
ideal lab to testify the hadron structures and as well help to
identify the quantum numbers for a newly observed resonances. Even
though, the reaction mechanism for radiative decays is free of
strong interaction (at tree level), the non-perturbative QCD effects
are still involved in the hadronization of emerging hadrons, as well
as the wavefunction of the parent hadron. To search for new physics
or possible exotic components of the hadron, one must thoroughly
study how the non-perturbative QCD affects the decay rates in a
reasonable theoretical framework. In this work, we are going to
derive the formulas for the radiative decays of $0^{++}$ and
$1^{+-}$ heavy mesons which are supposed to have regular $q\bar q'$
structures, in the light front quark model (LFQM). Through a
comparison of the theoretical predictions with data, the consistency
degree would reveal if the concerned hadron possesses the regular
$q\bar q'$ structure or has an exotic component.

During recent years several new particles such as
$D_s(2317)$\cite{Aubert:2003fg},
$D_s(2460)$\cite{Besson:2003cp,Krokovny:2003zq},
$X(3872)$\cite{Choi:2003ue}, $X(3940)$\cite{Abe:2007jn} and
$Y(3940)$\cite{Choi:2005} have been observed in experiments and
re-confirmed as new resonances. However it is not easy to identify
their inner structures, i.e. if the constituents of the resonances
are indeed just $q$ and $\bar q'$ or with something else (gluon or
$q\bar q$ pair etc.). Some of them are speculated as exotic states.
The reasons may be twofold: one is their peculiar decay modes
whereas the other is that the theoretical expectations on their
excited states are not well consistent with data or even missing.
For example when $D_s(2317)$ was announced it was considered as a
molecular state or a tetraquark, however it may just be a regular
$p-$wave $c\bar s$ meson with $J^{P}=0^{+}$ as it appears in
particle data book\cite{PDG12}. Because radiative decay is fully
governed by the electromagnetic interaction, the reaction mechanism
is relatively simple and the emitted photon can be well measured in
experiment, it may offer a good opportunity to justify the quantum
numbers and constituent structure of the newly observed resonance.
Namely, one may check whether with the simple $q\bar q'$ assignment
the theoretical prediction agrees with data.

For $p-$wave particles there are three degenerate states $0^{++}$,
$1^{++}$ and $2^{++}$ with the total intrinsic spin $S=1$ and one
singlet $1^{+-}$ with $S=0$. It is well known that  $\chi_{c0},
\chi_{c1}$ and $\chi_{c2}$ and $h_c$ are triplets and singlet
$p-$wave charmonia respectively.

In this work we will calculate the rates of $0^{++}\rightarrow
1^{--}\gamma$ and $1^{+-}\rightarrow 0^{-+}\gamma$ in LFQM
\cite{Cheng:1996if,Cheng:2003sm,Choi:2007se,Hwang:2006cua,Jaus,Ji:1992yf,Ke:2007tg,Li:2010bb}
which has been successfully applied to evaluate rates of
semileptonic and non-leptonic decays of $s-$wave heavy mesons. For
the $p-$wave mesons, the wave functions of $0^{++}$, $1^{--}$,
$1^{+-}$ and $0^{-+}$ have been constructed and the leading Feynman
diagrams are simple, with them we are able to calculate the
corresponding transition amplitudes.

Since the branching ratio of $\chi_{c0}\rightarrow J/\psi\gamma$ is
well measured, we first calculate $B(\chi_{c0}\rightarrow
J/\psi\gamma)$ in LFQM  to fix the model parameters and check the
validity degree of this approach, then with the formulas for the
transition $0^{++}\rightarrow 1^{--}\gamma$ we estimate the rates of
$\chi_{b0}\rightarrow \Upsilon(1S)\gamma$ and $D_s(2317)\rightarrow
D_s^*+\gamma$. These formulas can also be applied to study decays of
other $0^{++}$ states. Recently the spin singlets $h_c$ and $h_b$
attract intensive interests of both experimentalists and
theorists\cite{Li:2012rn,Fleming:1998md,Godfrey:2005un}. By the
formulas for $1^{+-}\rightarrow 0^{-+}\gamma$ we calculate the
widths of $h_c\rightarrow \eta_c\gamma$ and $h_b\rightarrow
\eta_b\gamma$. The results can be compared with the data which will
be available soon at the BES II and future B-factory. This
comparison definitely assures us if they are pure $p-$wave heavy
quarkonia.

After the introduction we derive the formulas for the transition
$0^{++}\rightarrow 1^{--}\gamma$ and $1^{+-}\rightarrow
0^{-+}\gamma$ in section II. Then in Sec. III, we numerically
evaluate the decay widths of $\chi_{c0}\rightarrow J/\psi\gamma$,
$\chi_{b0}\rightarrow \Upsilon(1S)\gamma$, $D_s(2317)\rightarrow
D_s^*+\gamma$, $h_c\rightarrow \eta_c\gamma$ and $h_b\rightarrow
\eta_b\gamma$ and make some discussions. In the last section we give
a brief summary and discussion. Some  notations and definitions of
relevant quantities are collected in the attached appendix.

\section{the formula for the decays $0^{++}\rightarrow 1^{--}\gamma$ and $1^{+-}\rightarrow 0^{-+}\gamma$}
In Ref.\cite{Chung:1993da} Chung studied various transitions and
derived the corresponding  amplitudes, for example, he determined
the amplitude structure of $b_1(1235)\rightarrow \omega\pi$. For
transition $1^{+-}\rightarrow 0^{-+}\gamma$ \cite{Chung:1993da} a
photon exists in the final states, thus the gauge invariance demands
the transition amplitude to be in the form
\begin{eqnarray}\label{p1}
A=F*(g_{\mu\nu}-\frac{k_\mu q_\nu}{k\cdot
q})\varepsilon^\mu\varepsilon'^\nu,
\end{eqnarray}
where $q$ and $k$ represent the momenta of the photon and daughter
meson, $\varepsilon'$ and $\varepsilon$ are the polarizations of the
vector-meson and photon respectively, and $F$ is the form factor
which is what we are going to derive and numerically compute in this
work. Though the $0^{++}\rightarrow 1^{--}1^{--}$ was not discussed
in Ref.\cite{Chung:1993da}, its amplitude structure is the same as
Eq.(\ref{p1}) and it can be seen from analyzing the $J^{P}$
characters of the involved mesons.

Below, following the schemes given in literature, we will calculate
the transition amplitudes of interest in LFQM.

\subsection{the decay of $0^{++}\rightarrow 1^{--}\gamma$}
The vertex functions of $0^{++}$ and $1^{--}$ are respectively
\cite{Cheng:2003sm}
\begin{eqnarray}
&& -iH_S\\
&& iH_V[\gamma_\mu-\frac{1}{W_V}(p_a-p_b)_\mu]
\end{eqnarray}
where $H_S$, $H_V$ and $W_V$ are defined in Ref.\cite{Cheng:2003sm}
and $p_a$ and $p_b$ are the momenta of constituents of the
corresponding meson.

\begin{figure}
\begin{center}
\begin{tabular}{ccc}
\scalebox{0.5}{\includegraphics{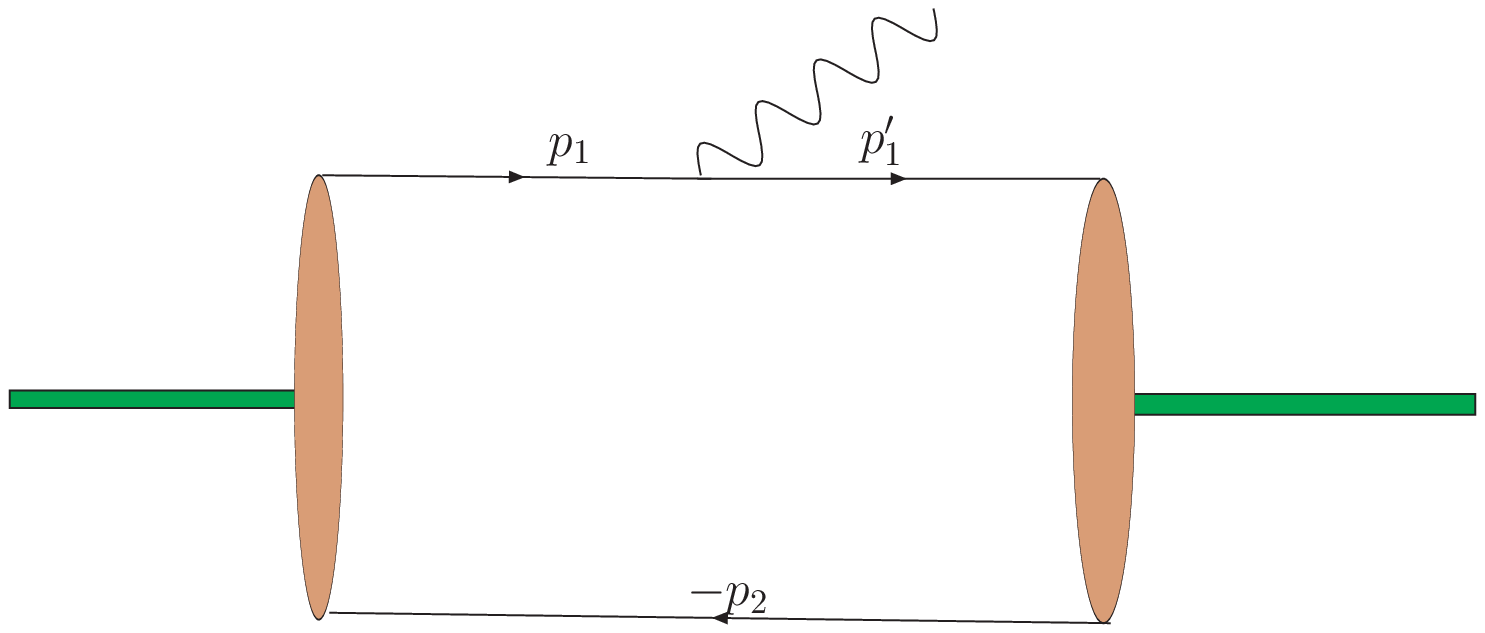}}\scalebox{0.5}{\includegraphics{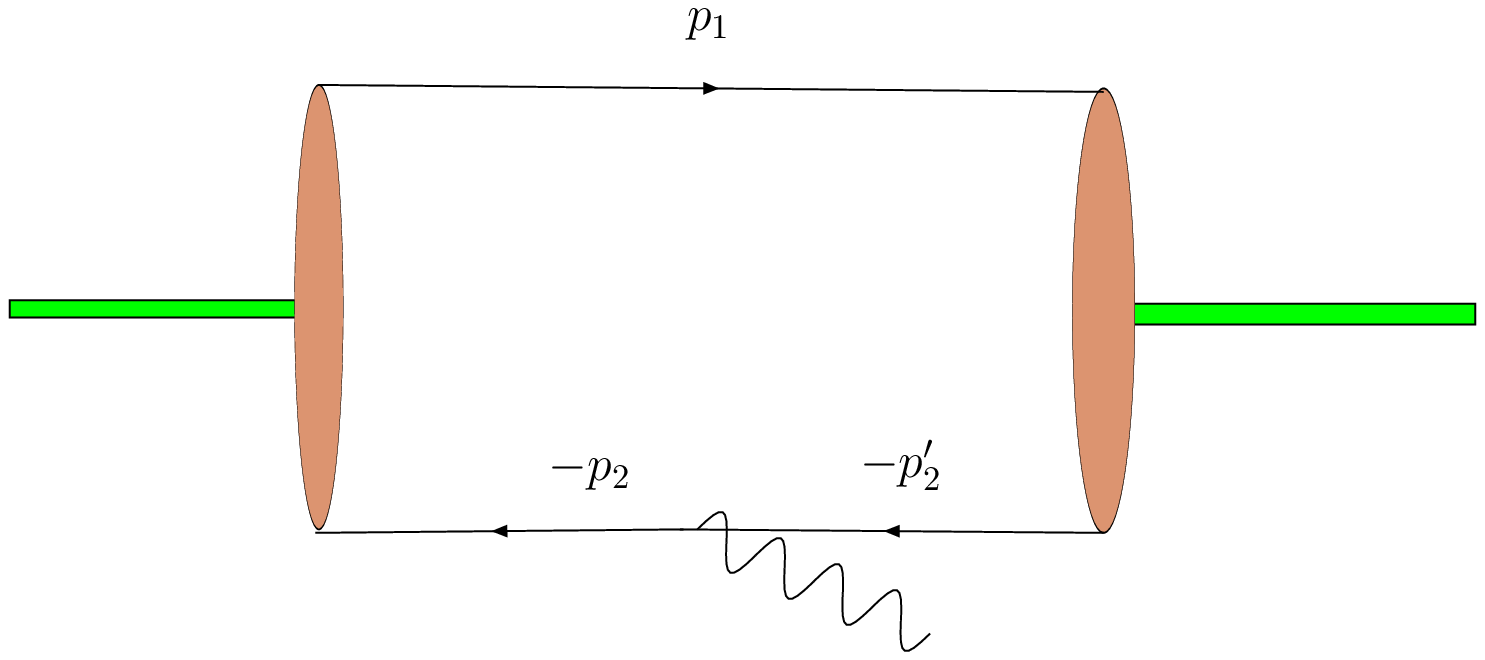}}
\end{tabular}
\end{center}
\caption{Feynman diagrams for radiative decay.}\label{t1}
\end{figure}

The  transition amplitude corresponding to the left diagram of Fig.1
is written as
\begin{eqnarray}
\mathcal{A}^a=iee_1\frac{N_c}{(2\pi)^4}\int d^4
p_1\frac{H_SH_V}{N_1N_1'N_2}s^a_{\mu\nu}\varepsilon'^\nu\varepsilon^\mu
\end{eqnarray}
where  $$s_{\mu\nu}^a={\rm
Tr}\{(-\slash\!\!\!p_2+m_2)[\gamma_\nu-\frac{(p'_1-p_2)_\nu}
{W_{V}}](\slash\!\!\!p'_1+m_1)\gamma_\mu(\slash\!\!\!p_1+m_1)\},$$
 $N_1=p_1^2-m_1^2+i\epsilon$,
$N_1'={p'}_1^2-m_1^2+i\epsilon$, $N_2=p_2^2-m_2^2+i\epsilon$,
$ee_{1(2)}$ is the electric charge of the quark of u- or d-types. In
the light front frame, $p_i$ is decomposed as
($p_i^-,p_i^+,{p_i}_\perp$). Integrating out $p_{1}^-$ with the
methods given in Ref.\cite{Cheng:1996if} one has
\begin{eqnarray}\label{vf9.2}
\int d^4p_1
\frac{H_{S}H_V}{N_1N_1'N_2}s_{\mu\nu}^a{\varepsilon'}^\nu
{\varepsilon}^{\mu}\rightarrow-i\pi\int
dx_1d^2p_\perp\frac{h_{S}h_V}{x_2 \hat{N_1}\hat{N_1'}}\hat
s_{\mu\nu}^a\hat{\varepsilon'}^\nu \hat{\varepsilon}^{\mu},
\end{eqnarray}
with
\begin{eqnarray*}
h_{S}=&&(M^2-M_0^2)\sqrt{\frac{x_1x_2}{N_c}}\frac{1}{\sqrt{2}\tilde{M_0}}\frac{\tilde{M_0}^2}{2\sqrt{3}M_0}\frac{\sqrt{2}}{\beta}\phi(nS),\\
h_V=&&({M'}^2-{M'}_0^2)\sqrt{\frac{x'_1x'_2}{N_c}}\frac{1}{\sqrt{2}\tilde{M'}_0}\phi'(nS),\\
\hat{N}_1^{(')}=&&x^{(')}_1({M^{(')}}^2-{M_0^{(')}}^2),
\end{eqnarray*}
where $M$ and $M'$ represent the masses of decaying and produced
mesons and the relation between $\varepsilon(\varepsilon')$ and
$\hat\varepsilon(\hat\varepsilon')$ can be found in the appendix
of Ref.\cite{Cheng:2003sm}.

To include the contributions from the zero mode, ${p_1}_\mu$,
${p_1}_\nu$, ${p_1}_\mu {p_1}_\nu$ and $W_V$ in $s_{\mu\nu}^a$
should be replaced by appropriate expressions as discussed in
Ref.\cite{Cheng:2003sm}, for example
\begin{eqnarray}\label{p6}
&&W_V\rightarrow w_V=M_0+m_1+m_2\nonumber\\
&&{p_1}_\mu\rightarrow\frac{x_1}{2}\mathcal{P}_\mu+(\frac{x}{2}-\frac{p_\perp
q_\perp}{q^2})q_\mu,\nonumber\\
&&\,\,\,\,\,\,\,\,\,\,\,\,\,\,......
\end{eqnarray}
with $\mathcal{P}=p+k$ and $p$ is the momentum of decaying meson.

The definitions of $M_0$ and  $\tilde{M}_0$ are presented in the
appendix, as more details about the derivations and relevant
notations can be found in Ref.\cite{Cheng:2003sm}. In this
framework, $\hat s_{\mu\nu}^a$ replaces $s_{\mu\nu}^a$ and is
written into the form
\begin{eqnarray}\label{p7}
\hat s_{\mu\nu}^a={F_1} g_{\mu\nu}+{F_2}\frac{k_\mu q_\nu}{k\cdot
q},
\end{eqnarray}
with
\begin{eqnarray}
{F_1}=&&-2\,\left[ {m_1}\,\left( {{M}}^2 - {{M'}}^2 - {\hat{N}_1}
+ {\hat{N}_1'} \right)  +
     {m_2}\,\left( {\hat{N}_1} + {\hat{N}_1'} - {q^2} \right)  \right]  -
   \nonumber\\&&\frac{4\,{A12}\,\left( 2\,{{m_1}}^2 + 4\,{m_1}\,{m_2} + 2\,{{m_2}}^2 - {{M}}^2 - {{M'}}^2 + {q^2} -
       2\,{m_1}\,{w_V} - 2\,{m_2}\,{w_V} + 2\,{Z2} \right) }{{w_V}},\nonumber\\
{F_2}=&&\frac{4\,k\cdot q\,{A_2^{(1)}}\,\left( -{\hat{N}_1} -
{\hat{N}_1'} + {q^2} \right) }{{w_V}} +
  \frac{4\,k\cdot q\,\left( {\hat{N}_1} + {\hat{N}_1'} - {q^2} + {m_1}\,{w_V} \right) }{{w_V}} +
 \nonumber\\&& \frac{4\,k\cdot q\,{A_2^{(2)}}\,\left\{-2\,\left[ 2\,{\left( {m_1} + {m_2} \right) }^2 - {{M}}^2 - {{M'}}^2 + {q^2} \right]  +
       4\,{m_1}\,{w_V} + 4\,{m_2}\,{w_V} \right\} }{{w_V}} +
 \nonumber\\&&  \frac{4\,k\cdot q\,{A_3^{(2)}}\,\left\{ -2\,\left[ 2\,{\left( {m_1} + {m_2} \right) }^2 - {{M}}^2 - {{M'}}^2 + {q^2} \right]  +
       4\,{m_1}\,{w_V} + 4\,{m_2}\,{w_V} \right\} }{{w_V}} +
 \nonumber\\&&  \frac{4\,k\cdot q\,{A_1^{(1)}}\,\left[ 4\,{\left( {m_1} + {m_2} \right) }^2 - 2\,{{M}}^2 - 2\,{{M'}}^2 - {\hat{N}_1} - {\hat{N}_1'} +
       3\,{q^2} - 2\,\left( 3\,{m_1} + {m_2} \right) \,{w_V} \right] }{{w_V}},
\end{eqnarray}
and $A_{i}{^{(j)}}(i=1\sim4,j=1\sim4)$ are presented in the attached
appendix.

We define the form factors as following
\begin{eqnarray}
&&{f_1}(m_1,m_2,e_1)=\frac{ee_1}{32\pi^3}\int dx_2d^2p_\perp \frac{{F_1}\phi\phi'\tilde{M_0}}{\sqrt{6}\beta M_0\tilde{M'}_0x_1},\nonumber\\
&&{f_2}(m_1,m_2,e_1)=\frac{ee_1}{32\pi^3}\int dx_2d^2p_\perp
\frac{{F_2}\phi\phi'\tilde{M_0}}{\sqrt{6}\beta
0M_0\tilde{M'}_0x_1},
\end{eqnarray}
where $\phi$ and $\phi'$ are respectively the wavefunctions of the
initial and final mesons. These form factors will be numerically
evaluated in next section.

With these form factors   the amplitude corresponding to the left
diagram of Fig.1 is obtained as
\begin{eqnarray}\label{vf9.1}
\mathcal{A}^a&&= {f_1}(m_1,m_2,e_1)
\varepsilon\cdot\varepsilon'+{f_2}(m_1,m_2,e_1)
\frac{k\cdot\varepsilon' q\cdot\varepsilon}{k\cdot q}.
\end{eqnarray}

The right diagram is just the charge conjugation of the left one of
Fig.1, so that one can immediately write it down
\begin{eqnarray}\label{vf9.1}
\mathcal{A}^b&&= {f_1}(m_2,m_1,e_2)
\varepsilon\cdot\varepsilon'+{f_2}(m_2,m_1,e_2)
\frac{k\cdot\varepsilon' q\cdot\varepsilon}{k\cdot q}.
\end{eqnarray}

The total amplitude is simply a sum of $\mathcal{A}^a$ and
$\mathcal{A}^b$:
\begin{eqnarray}\label{vf9.1}
\mathcal{A}&&=\mathcal{A}^a+\mathcal{A}^b.
\end{eqnarray}

Comparing Eq. (1) with Eq. (12) we determine the full form factor as
\begin{eqnarray}\label{c1}
F=f_1(m_1,m_2,e_1)+f_1(m_2,m_1,e_2)=-[f_2(m_1,m_2,e_1)+f_2(m_2,m_1,e_2)].
\end{eqnarray}
In principle one can calculate
$f_1=f_1(m_1,m_2,e_1)+f_1(m_2,m_1,e_2)$ and
$f_2=f_2(m_1,m_2,e_1)+f_2(m_2,m_1,e_2)$ separately and fix the form
factor, then go on obtaining the decay rate.

\subsection{the decay rate of $1^{+-}\rightarrow 0^{-+}\gamma$}
The vertex functions of $0^{-+}$ and $1^{+-}$ are presented in
Ref.\cite{Cheng:2003sm}, they are respectively
\begin{eqnarray}
&& H_P\gamma_5;\\
&& -iH_{^1A}[\frac{1}{W_{^1A}}(p_a-p_b)_\mu]\gamma_5,
\end{eqnarray}
where $H_P$, $H_{^1A}$ and $W_{^1A}$ are some relevant functions
which are slightly lengthy and can be found in
Ref.\cite{Cheng:2003sm}, for saving space, we do not repeat them
here.

The  transition amplitude corresponding to the left diagram of
Fig.1 is written as
\begin{eqnarray}
\mathcal{A}^a=ee_1\frac{N_c}{(2\pi)^4}\int d^4
p_1\frac{H_PH_{^1A}}{N_1N_1'N_2}s^a_{\mu\nu}\varepsilon'^\nu\varepsilon^\mu
\end{eqnarray}
where  $$s_{\mu\nu}^a={\rm
Tr}\{(-\slash\!\!\!p_2+m_2)[\frac{(p'_1-p_2)_\nu}
{W_{^1A}}\gamma_5](\slash\!\!\!p'_1+m_1)\gamma_\mu(\slash\!\!\!p_1+m_1)\gamma_5\}.$$
Integrating out $p_{1}^-$  we have
\begin{eqnarray}\label{vf9.2}
\int d^4p_1
\frac{H_{P}H_{^1A}}{N_1N_1'N_2}s_{\mu\nu}^a{\epsilon'}^\nu
{\epsilon}^{\mu}\rightarrow-i\pi\int
dx_1d^2p_\perp\frac{h_{P}h_{^1A}}{x_2 \hat{N_1}\hat{N_1'}}\hat
s_{\mu\nu}^a\hat{\varepsilon'}^\nu \hat{\varepsilon}^{\mu},
\end{eqnarray}
where
\begin{eqnarray*}
h_{{^1A}}=&&(M^2-M_0^2)\sqrt{\frac{x_1x_2}{N_c}}\frac{1}{\sqrt{2}\tilde{M_0}}\frac{\sqrt{2}}{\beta}\phi(nS),\\
h_P=&&({M'}^2-{M'}_0^2)\sqrt{\frac{x'_1x'_2}{N_c}}\frac{1}{\sqrt{2}\tilde{M'}_0}\phi'(nS).
\end{eqnarray*}

With a replacement similar to Eq.(\ref{p6}), $\hat s_{\mu\nu}^a$ is
re-written into the form of Eq. (\ref{p7}) with
\begin{eqnarray}
{F_1}=&&{A_1^{(2)}}\,( \frac{-8\,{{m_1}}^2}{{w_{^1A}}} +
\frac{16\,{m_1}\,{m_2}}{{w_{^1A}}} -
\frac{8\,{{m_2}}^2}{{w_{^1A}}} +
    \frac{4\,{{M}}^2}{{w_{^1A}}} + \frac{4\,{{M'}}^2}{{w_{^1A}}} - \frac{4\,q^2}{{w_{^1A}}} - \frac{8\,{Z_2}}{{w_{^1A}}} )
\nonumber\\
{F_2}=&&{2\,k\cdot q [A_1^{(1)}}\,(
\frac{2\,{\hat{N}_1}}{{w_{^1A}}} +
\frac{2\,{\hat{N}_1'}}{{w_{^1A}}} - \frac{2\,q^2}{{w_{^1A}}} ) +
  {A_2^{(1)}}\,( \frac{-2\,{\hat{N}_1}}{{w_{^1A}}} - \frac{2\,{\hat{N}_1'}}{{w_{^1A}}} + \frac{2\,q^2}{{w_{^1A}}} )  +
 \nonumber\\&&  {A_2^{(2)}}\,( \frac{8\,{{m_1}}^2}{{w_{^1A}}} - \frac{16\,{m_1}\,{m_2}}{{w_{^1A}}} + \frac{8\,{{m_2}}^2}{{w_{^1A}}} -
     \frac{4\,{{M}}^2}{{w_{^1A}}} - \frac{4\,{{M'}}^2}{{w_{^1A}}} + \frac{4\,q^2}{{w_{^1A}}} )
+ \nonumber\\&&{A_3^{(2)}}\,( \frac{-8\,{{m_1}}^2}{{w_{^1A}}} +
\frac{16\,{m_1}\,{m_2}}{{w_{^1A}}} -
\frac{8\,{{m_2}}^2}{{w_{^1A}}} +
     \frac{4\,{{M}}^2}{{w_{^1A}}} + \frac{4\,{{M'}}^2}{{w_{^1A}}} - \frac{4\,q^2}{{w_{^1A}}} )]  .
\end{eqnarray}

\section{applying the approach to analyze radiative decays}

In this section we first test the formula by comparing our
theoretical evaluation on the rate of a well measured decay mode
with the data and confirm its validity, then apply it to predict
the rates of radiative decays of $0^{++}$ and $1^{+-}$ which will
be experimentally measured soon. We select transition
$\chi_{c0}\rightarrow J/\psi\gamma$  as a probe to check the
approach since its branching ratio is well measured. Setting
$m_c=1.4$GeV and the model parameters
$\beta_{\psi(\chi_c)}=0.631$GeV\cite{Ke:2011jf}\footnote{We vary
the parameters within a $\pm$10\% range to estimate the
corresponding errors.}, we get the form factors $f_1$ and $f_2$
for the transition $\chi_{c0}\rightarrow J/\psi\gamma$ which are
presented in Tab. I. One can immediately notice that $f_1$ is
sensitive to the variation of the parameters, especially $m_c$,
but $f_2$ is insensitive. The problem originates from the fact
that $\hat{N_1^{(')}}$ is proportional to a cancelation of two
large numbers. Even though $\hat{N_1^{(')}}$ resides in both $F_1$
and $F_2$, its contribution is suppressed by $w_V$ or $w_{^1A}$ in
$F_2$, so that $F_2$ is insensitive to the parameters. For
reducing uncertainties of our theoretical computations, we will
calculate $F$ in terms of the relation between $f_2$ and $F$ in
Eq.(\ref{c1}). Our theoretical estimate of the width of
$\chi_{c0}\rightarrow J/\psi\gamma$ is (152$\pm$31)keV and the
corresponding branching ratio is (1.46$\pm$0.31)\% which is
consistent with the data $(1.17\pm0.08)$\%\cite{PDG12} within a
tolerable error range.

Then we study the transition $\chi_{b0}\rightarrow
\Upsilon(1S)\gamma$ using  the parameter $m_b=4.8$GeV and
$\beta_{\Upsilon(\chi_b)}=1.288$GeV which were fixed in
Ref.\cite{Ke:2010x} by fitting other well measured channels. Our
estimate of $\Gamma(\chi_{b0}\rightarrow \Upsilon(1S)\gamma)$ is
(21.3$\pm$4.7)keV. With the branching ratio
(1.76$\pm$0.35)\%\cite{PDG12},  the total width of $\chi_{b0}$   is
estimated to be  (1.21$\pm$0.36) MeV.

\begin{table}
\caption{the form factors $f_1$ and $f_2$} \label{tab:decay}
\begin{tabular}{c|c|c|c|c|c}\hline\hline
 decay mode   &  ~~~~~~$f_1$~~~~~~   &
 ~~~~~~$f_2$~~~~~~  & width(keV) & $\mathcal{BR}$(the)& $\mathcal{BR}$(exp) \\\hline
 $\chi_{c0}\rightarrow J/\psi\gamma$  & -2.83 $\pm$1.37  & 0.89$\pm$0.09&152$\pm$31 & (1.46$\pm$0.31)\%     & $(1.17\pm0.08)$\% \cite{PDG12}      \\
$\chi_{b0}\rightarrow \Upsilon(1S)\gamma$ & -0.50$\pm$3.01   &
0.82$\pm$0.08&21.3$\pm$4.7 &  - &(1.76$\pm$0.35)\%\cite{PDG12}
\\$D_s(2317)\rightarrow
{D_s}^*\gamma$ & -1.11$\pm$0.22   &0.25$\pm$0.03&17.1$\pm$3.9&$>(0.45\pm 0.11)\%$ &-\\
$h_c\rightarrow \eta_c\gamma$ & -2.36$\pm$0.34  &2.47$\pm$0.18&685$\pm$89&$>(69\pm 9)\%$ &$(51\pm 6)\%$\cite{PDG12} \\
$h_b\rightarrow \eta_b\gamma$ & -2.40$\pm$0.31  &1.78$\pm$0.20&36.9$\pm$8.7&- &  $(49.2\pm 5.7^{+5.6}_{-3.3})\% $\cite{Mizuk:2012pb}  \\
\hline\hline
\end{tabular}
\end{table}

\begin{table}
\caption{{Predictions made in terms of various approaches and
experimental data (if available) (in units of keV)}}
\label{tab:com}
\begin{tabular}{c|c|c|c|c}\hline\hline
 decay mode   &  ~~~~~~\cite{Ebert:2002pp}~~~~~~   &
 ~~~~~~\cite{DeFazio:2008xq}~~~~~~  &our results & exp \\\hline
 $\chi_{c0}\rightarrow J/\psi\gamma$  &121&   input  &152$\pm$31  &122$\pm$12 \cite{PDG12}      \\
$\chi_{b0}\rightarrow \Upsilon(1S)\gamma$ &29.9  &
85$\pm$4&21.3$\pm$4.7 &-
\\
$h_c\rightarrow \eta_c\gamma$ & 560&634$\pm$32&685$\pm$89&-\\
$h_b\rightarrow \eta_b\gamma$ & 52.6  &271$\pm$14&36.9$\pm$8.7 &  - \\
\hline\hline
\end{tabular}
\end{table}


Supposing $D_s(2317)$ to be a regular bound state of $c\bar s$ and
setting $m_s=0.37$GeV, $\beta_{D_s}=0.592$GeV\cite{Wei:2009nc}, we
can calculate the rate of $D_s(2317)\rightarrow D_s^*+\gamma$ using
the above formulas. We obtain its partial width as
(17.1$\pm$3.9)keV. Namely, if $D_s(2317)$ is a $0^{++}$ regular
meson, possessing the $p-$wave $c\bar s$ structure, our numerical
prediction should be consistent with the data which will be
available at BES II or B-factory soon, therefore the consistency
degree with data can help to confirm or negate its assignment $c\bar
s$.

In fact these formulas can be applied to study radiative decays of
other heavy mesons such $D_0$, $B_{0}$, $B_{s0}$ and $B_{c0}$.

Once $h_c$ and $h_b$  were experimentally measured, their special
behaviors draw intensive interests of theorists. The main focus is
if they are regular quarkonia or have exotic components.  Using
the formulas for $1^{+-}\rightarrow0^{-+}\gamma$, we calculate the
branching ratios of $h_c\rightarrow \eta_c\gamma$ and
$h_b\rightarrow \eta_b\gamma$ and the numerical result are
presented in Tab. I, which can be used to analyze their
characters. In terms of the measured upper limit of the total
width of $h_c$ we estimate the branching ratio of
$\mathcal{B}(h_c\rightarrow \eta_c\gamma)>(69\pm9)\%$ which is a
little larger than  the data $(51\pm 6)\%$. If  $h_c$ ($h_b$) is a
pure charmonium (bottomonium) our estimate indicates that its
total width should be around 1 MeV (80keV). To further identify
the inner structures of $h_c$ and $h_b$, more precise experiments
are needed.

It is noted that the masses of initial and finial mesons we used
in our numerical computations are taken from Ref.\cite{PDG12}
except the  mass of $\eta_b$ which is
9402.4MeV\cite{Mizuk:2012pb}.  Some parallel researches on these
decays can be found in Ref.\cite{DeFazio:2008xq,Ebert:2002pp} {and
a comparison of  their results with ours is made and listed in
Tab. \ref{tab:com}. Because of large errors in the inputs for our
numerical computations, our predictions are of relatively large
uncertainties. To further testify the model or constrain the model
parameters, much more precise experiments are needed.  }

\section{summary}
In this work we derive the formulas for the radiative decay of
$0^{++}$ and $1^{+-}$ heavy mesons and numerically compute the
rates. We formulate the transition matrix elements and extract the
form factors in the LFQM. To check the validity degree of this
approach where the model parameters were fixed by fitting other
physical processes in previous works, we calculate the decay
widths of $\chi_{c0}\rightarrow J/\psi\gamma$ and compare it with
data. Considering both theoretical and measurement uncertainties,
the theoretical result on the branching ratio of
$\chi_{c0}\rightarrow J/\psi\gamma$ (1.46$\pm$0.31)\% is
satisfactorily consistent with data $(1.17\pm0.08)\%$. With the
same scenario, we predict the width of $\chi_{b0}\rightarrow
\Upsilon(1S)\gamma$ to be about (21.3$\pm$4.7)keV and the width of
$D_s(2317)\rightarrow D_s^*+\gamma$ to be (17.1$\pm$3.9)keV.
Comparing those results with the data which will be available at
BES and B-factory, one can further confirm the inner structure of
$\chi_{b0}$ and $D_s(2317)$.

In terms of the data $B(\chi_{b0}\rightarrow \Upsilon(1S)\gamma)$
we estimate the total width of $\chi_{b0}$ as (1.21$\pm$0.36)MeV
which would be easier to be experimentally checked. As for the
$1^{+-}$ decay, we evaluate the decay rates of $h_c\rightarrow
\eta_c\gamma$ and $h_b\rightarrow \eta_b\gamma$ which are
$(685\pm89)$keV and $(36.9\pm8.7)$ keV respectively. These
formulas deduced in this work can also be applied to study
radiative decays of other $0^{++}$ and $1^{+-}$ particles.

\section*{Acknowledgement}

This work is supported by the National Natural Science Foundation
of China (NNSFC) under the contract No. 11075079 and No. 11005079;
the Special Grant for the Ph.D. program of Ministry of Eduction of
P.R. China No. 20100032120065.

\appendix
\section{Model description}
{In the conventional light-front model, a meson containing a quark
$q_1$ and an antiquark $\bar q_2$ with its  total momentum $P$ and
spin  $J$ can be expressed\cite{Cheng:1996if}
\begin{eqnarray}\label{eq:lfmeson}
 |\mathcal{M}(P^{2S+1},L_J,J_z)\rangle&=&\int\{d^3p_1\}\{d^3p_2\} \,
  2(2\pi)^3\delta^3(\tilde{P}-\tilde{p_1}-\tilde{p_2}) \nonumber\\
 &&\times\sum_{\lambda_1,\lambda_2}\Psi^{JJ_z}_{LS}(\tilde{p}_1,\tilde{p}_2,\lambda_1,\lambda_2)
 \left|\right.
 q_1(p_1,\lambda_1)\bar q_2(p_2,\lambda_2)\ra,
\end{eqnarray}
where
$\Psi^{JJ_z}_{LS}(\tilde{p}_1,\tilde{p}_2,\lambda_1,\lambda_2)$ is
wave function in momentum-space,  $\lambda_1$ and $\lambda_2$ denote
helicities, $p_1,~ p_2$ are the on-mass-shell light-front momenta
defined by
\begin{equation}
 \tilde{p_i}=(p_i^+,p_{i\perp}),\qquad p_{i\perp}=(p_i^1,p_i^2),\qquad
 p^-_{i}=\frac{m^2+p_{i\perp}^2}{p^+_{i}},
\end{equation}
and
\begin{eqnarray}
&&\{d^3p_i\}\equiv\frac{dp^+_{i}d^2 p_{i\perp}}{2(2\pi)^3},\qquad
  \delta^3(\tilde{p_i})=\delta(p^+_{i})\delta^2(p_{i\perp}).
\end{eqnarray}}

{From the eigen-equation\cite{Cheung:1995ub}
\begin{eqnarray}
\mathcal{H}_{LF}|\mathcal{M}(P^{2S+1},L_J,J_z)\rangle=\frac{M^2+P_{\perp}^2}{P^+_{i}}|\mathcal{M}(P^{2S+1},L_J,J_z)\rangle,
\end{eqnarray}
one can deduce a light-front Bethe-Salpeter equation
\begin{eqnarray}\label{bse}
(M^2-M_0^2)\phi(x,p_\perp)=\int\frac{dx'd^2p_\perp'}{2(2\pi)^3}V_{eff}
\phi(x',p_\perp').
\end{eqnarray}
In principle by solving Eq.(\ref{bse}) one can obtain the
momentum distribution amplitude $\phi(x,p_\perp)$. However the
two-body interaction kernel $V_{eff}$ is complicate so some
phenomenological amplitudes are chosen in practicable theoretical calculations. In this work we use the
Gaussian-type amplitudes, for example
\begin{eqnarray}
\phi(1S)=4(\frac{\pi}{\beta^2})^{3/4}\sqrt{\frac{dp_z}{dx_2}}{\rm
 exp}(-\frac{p^2_z+p^2_\perp}{2\beta^2}),
 \end{eqnarray}
where  $\beta$ is the model parameter to be fixed by
calculating the decay constant. A parallel approach about LFQM
was done in Refs.\cite{Choi:1997iq,Choi:1999nu} where the Gaussian-type amplitude is adopted as a trial
function for the variational computations of the QCD-motivated
effective Hamiltonian.}
\section{Notations}

Here we list some variables and notations appearing in the context.
The incoming meson in Fig. \ref{t1} has momentum $P=p_1+p_2$
where $p_1$ and $p_2$ are the momenta of the off-shell quark and
antiquark and
\begin{eqnarray}\label{app1}
&& p_1^+=x_1P^+, \qquad ~~~~~~p_2^+=x_2P^+, \nonumber\\
&& p_{1\perp}=x_1P_{\perp}+p_\perp, \qquad
 p_{2\perp}=x_2P_{\perp}-p_\perp,
 \end{eqnarray}
with  $x_i$ and $p_\perp$ are internal variables and $x_1+x_2=1$.

The variables $M_0$, $\tilde {M_0}$ and $\hat{N_1}$ are defined as
\begin{eqnarray}\label{app2}
&&M_0^2=\frac{p^2_\perp+m^2_1}{x_1}+\frac{p^2_\perp+m^2_2}{x_2},\nonumber\\&&
\tilde {M_0}=\sqrt{M_0^2-(m_1-m_2)^2}.
 \end{eqnarray}
with $p_z=\frac{x_2M_0}{2}-\frac{m_2^2+p^2_\perp}{2x_2M_0}$.

$A_{ij}(i=1\sim 4, j=1\sim4)$ are
\begin{eqnarray}\label{app2}
&&A_{1}{^{(1)}}=\frac{x_1}{2},\,\,\,
A_{2}{^{(1)}}=A_{1}{^{(1)}}-\frac{p_{\perp}\cdot
q_{\perp}}{q^2},\,\,\,A_{1}{^{(2)}}=-p_{\perp}^2-\frac{(p_{\perp}\cdot
q_{\perp})^2}{q^2},\nonumber\\&&
A_{2}{^{(2)}}=({A_{1}{^{(1)}}})^2,\,\,\,A_{3}{^{(2)}}=A_{1}{^{(2)}}A_{2}{^{(2)}},\,\,\,A_{4}{^{(2)}}=(A_{2}{^{(1)}})^2-\frac{A_{1}{^{(2)}}}{q^2},
\nonumber\\&&A_{1}{^{(3)}}=A_{1}{^{(1)}}A_{12},\,\,\,A_{2}{^{(3)}}=A_{2}{^{(1)}}A_{1}{^{(2)}},\,\,\,
A_{3}{^{(3)}}=A_{1}{^{(1)}}A_{2}{^{(2)}},\,\,\,\nonumber\\&&A_{4}{^{(3)}}=A_{2}{^{(1)}}A_{2}{^{(2)}},
A_{1}{^{(4)}}=\frac{(A_{1}{^{(2)}})^2}{3},\,\,\,A_{2}{^{(4)}}=A_{1}{^{(1)}}A_{1}{^{(3)}},\,\,\,\nonumber\\&&
A_{3}{^{(4)}}=A_{1}{^{(1)}}A_{2}{^{(3)}},\,\,\,A_{4}{^{(4)}}=A_{2}{^{(1)}}
A_{1}{^{(3)}}-\frac{A_{1}{^{(4)}}}{q^2}.
 \end{eqnarray}


\begin{thebibliography}{99}
\bibitem{Aubert:2003fg}
  B.~Aubert {\it et al.}  [BABAR Collaboration],
   Phys.\ Rev.\ Lett.\  {\bf 90}, 242001 (2003)  [hep-ex/0304021].  

\bibitem{Besson:2003cp}
  D.~Besson {\it et al.}  [CLEO Collaboration],
   Phys.\ Rev.\ D {\bf 68}, 032002 (2003)  [Erratum-ibid.\ D {\bf 75}, 119908 (2007)]  [hep-ex/0305100].  
\bibitem{Krokovny:2003zq}
  P.~Krokovny {\it et al.}  [Belle Collaboration],
  Phys.\ Rev.\ Lett.\  {\bf 91}, 262002 (2003)  [hep-ex/0308019].  


\bibitem{Choi:2003ue}
  S.~K.~Choi {\it et al.}  [Belle Collaboration],
  Phys.\ Rev.\ Lett.\  {\bf 91}, 262001 (2003)
  [arXiv:hep-ex/0309032].

\bibitem{Abe:2007jn}
  K.~Abe {\it et al.}  [Belle Collaboration],
  Phys.\ Rev.\ Lett.\  {\bf 98}, 082001 (2007)
  [arXiv:hep-ex/0507019].


\bibitem{Choi:2005} S.~K.~Choi {\it et al.}  [Belle Collaboration],
 Phys.\ Rev.\ Lett.\  {\bf 94}, 182002 (2005).



\bibitem{PDG12}
  J.~Beringer {\it et al.}  [Particle Data Group Collaboration],
  Phys.\ Rev.\ D {\bf 86}, 010001 (2012).  




\bibitem{Ke:2007tg}
  H.~W.~Ke, X.~Q.~Li and Z.~T.~Wei,
  Phys.\ Rev.\  D {\bf 77}, 014020 (2008)
  [arXiv:0710.1927 [hep-ph]];
  Z.~T.~Wei, H.~W.~Ke and X.~Q.~Li,
  Phys.\ Rev.\  D {\bf 80}, 094016 (2009)
  [arXiv:0909.0100 [hep-ph]];
  H.~W.~Ke, X.~Q.~Li and Z.~T.~Wei,
  Phys.\ Rev.\  D {\bf 80}, 074030 (2009)
  [arXiv:0907.5465 [hep-ph]];

%
  H.~W.~Ke, X.~Q.~Li and Z.~T.~Wei,
  Eur.\ Phys.\ J.\  C {\bf 69}, 133 (2010)
  [arXiv:0912.4094 [hep-ph]];
  H.~W.~Ke, X.~H.~Yuan and X.~Q.~Li,
Int. J. Mod. Phys. A {\bf 26}, 4731 (2010),
  arXiv:1101.3407 [hep-ph].
  H.~W.~Ke and X.~Q.~Li,
  Eur.\ Phys.\ J.\  C {\bf 71}, 1776 (2011)
  [arXiv:1104.3996 [hep-ph]];

  H.~W.~Ke and X.~Q.~Li,
  Eur.\ Phys.\ J.\  C {\bf 71}, 1776 (2011)
  [arXiv:1104.3996 [hep-ph]].



\bibitem{Jaus} W. Jaus, Phys. Rev.  D {\bf 41}, 3394 (1990);
  D {\bf 44}, 2851 (1991);  W.~Jaus,
  Phys.\ Rev.\  D {\bf 60}, 054026 (1999).

\bibitem{Ji:1992yf}
  C.~R.~Ji, P.~L.~Chung and S.~R.~Cotanch,
  Phys.\ Rev.\  D {\bf 45}, 4214 (1992).
\bibitem{Cheng:1996if}
  H.~Y.~Cheng, C.~Y.~Cheung and C.~W.~Hwang,
  Phys.\ Rev.\  D {\bf 55}, 1559 (1997)
  [arXiv:hep-ph/9607332].



\bibitem{Cheng:2003sm}
  H.~Y.~Cheng, C.~K.~Chua and C.~W.~Hwang,
  Phys.\ Rev.\  D {\bf 69}, 074025 (2004).


\bibitem{Hwang:2006cua}
  C.~W.~Hwang and Z.~T.~Wei,
  J.\ Phys.\ G {\bf 34}, 687 (2007);
%
  C.~D.~Lu, W.~Wang and Z.~T.~Wei,
  Phys.\ Rev.\  D {\bf 76}, 014013 (2007)
  [arXiv:hep-ph/0701265].


\bibitem{Choi:2007se}
  H.~M.~Choi,
  Phys.\ Rev.\  D {\bf 75}, 073016 (2007)
  [arXiv:hep-ph/0701263];


\bibitem{Li:2010bb}
  G.~Li, F.~l.~Shao and W.~Wang,
  Phys.\ Rev.\  D {\bf 82}, 094031 (2010)
  [arXiv:1008.3696 [hep-ph]].




\bibitem{Li:2012rn}
  J.~-Z.~Li, Y.~-Q.~Ma and K.~-T.~Chao,
  arXiv:1209.4011 [hep-ph].  
\bibitem{Fleming:1998md}
  S.~Fleming and T.~Mehen,
  Phys.\ Rev.\ D {\bf 58}, 037503 (1998)  [hep-ph/9801328].  

\bibitem{Godfrey:2005un}
  S.~Godfrey,
  ``Production of the h(c) and h(b) and implications for quarkonium spectroscopy,''  J.\ Phys.\ Conf.\ Ser.\  {\bf 9}, 123 (2005)  [hep-ph/0501083].



\bibitem{Chung:1993da}
  S.~U.~Chung,
  Phys.\ Rev.\ D {\bf 48}, 1225 (1993)  [Erratum-ibid.\ D {\bf 56}, 4419 (1997)].

\bibitem{Mizuk:2012pb}
  R.~Mizuk {\it et al.}  [Belle Collaboration],
  Phys.\ Rev.\ Lett.\  {\bf 109}, 232002 (2012)
  [arXiv:1205.6351 [hep-ex]].


\bibitem{Ke:2011jf}
  H.~W.~Ke and X.~Q.~Li,
  Phys.\ Rev.\  D {\bf 84}, 114026 (2011)
  [arXiv:1107.0443 [hep-ph]];

\bibitem{Ke:2010x}
  H.~W.~Ke, X.~Q.~Li, Z.~T.~Wei and X.~Liu,
  Phys.\ Rev.\  D {\bf 82}, 034023 (2010)
  [arXiv:1006.1091 [hep-ph]];



\bibitem{Wei:2009nc}
  Z.~T.~Wei, H.~W.~Ke and X.~F.~Yang,
  Phys.\ Rev.\  D {\bf 80}, 015022 (2009)
  [arXiv:0905.3069 [hep-ph]];







\bibitem{Ebert:2002pp}
  D.~Ebert, R.~N.~Faustov and V.~O.~Galkin,
  Phys.\ Rev.\ D {\bf 67}, 014027 (2003)
  [hep-ph/0210381].


\bibitem{DeFazio:2008xq}
  F.~De Fazio,
  Phys.\ Rev.\ D {\bf 79}, 054015 (2009)
  [Erratum-ibid.\ D {\bf 83}, 099901 (2011)]
  [arXiv:0812.0716 [hep-ph]];
  P.~Colangelo, F.~De Fazio and A.~Ozpineci,
  Phys.\ Rev.\ D {\bf 72}, 074004 (2005)
  [hep-ph/0505195].

\bibitem{Cheung:1995ub}
  C.~-Y.~Cheung, W.~-M.~Zhang and G.~-L.~Lin,
  Phys.\ Rev.\ D {\bf 52}, 2915 (1995)
  [hep-ph/9505232].

\bibitem{Choi:1997iq}
  H.~-M.~Choi and C.~-R.~Ji,
  Phys.\ Rev.\ D {\bf 59}, 074015 (1999)
  [hep-ph/9711450].
\bibitem{Choi:1999nu}
  H.~-M.~Choi and C.~-R.~Ji,
  Phys.\ Lett.\ B {\bf 460}, 461 (1999)
  [hep-ph/9903496].

\end{thebibliography}
\end{document}